# Ion-Implanted Erbium in X-cut Thin-film Lithium Niobate: Luminescence and Low-Temperature Response

*Daniel Blight\*, Mason Adshead, Alessandro Prencipe, Mayam Sanaee, Katia Gallo and Richard J Curry*

Dr. D. Blight, Dr. M. Adshead, Prof. R. J. Curry
Address: Photon Science Institute, Department of Electrical and Electronic Engineering, University of Manchester, Manchester, M13 9PL
Email Address: daniel.blight@manchester.ac.uk

Dr. A. Prencipe, Dr. M. Sanaee, Prof. K. Gallo
Address: Albanova University Center, Department of Physics, KTH – Royal Institute of Technology, Roslagstullsbacken 21, 10691 Stockholm, Sweden
Email Address: gallo@kth.se

Thin-film lithium niobate (TFLN) has emerged as a promising platform for photonic integrated circuits due to the strong electro-optic, piezoelectric and nonlinear properties unique to this well-established and technologically important crystal for electronic and photonic applications. Incorporation of rare-earth ions, particularly erbium ($Er^{3+}$), enables functionalities such as quantum memory, amplification, and single-photon emission in the telecom band. This study presents a method for deterministic $Er^{3+}$ doping of x-cut TFLN using focused ion beam (FIB) implantation with sub-100 nm spatial precision, enabling seamless integration of active rare-earth ions into this technologically relevant platform for lithium niobate integrated nanophotonics. Photoluminescence (PL) measurements from implanted regions reveal Stark-split 4f-4f transitions consistent with bulk Er-doped lithium niobate, indicating similar lattice occupation. Temperature-dependent PL measurements from 300 K to 5 K exhibit conventional behaviour down to approximately 50 K, followed by a marked decrease in the emission intensity and lifetime. This anomaly is attributed to a suppression of the pyroelectric response in $LiNbO_3$ at low temperatures, which affects local electric fields and, consequently, $Er^{3+}$ emission. The sensitivity of the PL response to the modulation frequency and polarization of the 980 nm excitation light is also consistent with possible mechanisms linking thermal effects and internal fields arising in the thin film. The results demonstrate a method for the targeted doping with $Er^{3+}$ ions into the most widely used cut of TFLN for integrated photonic devices and provide further important considerations for their exploitation in cryogenic quantum devices. The work paves the way for the deterministic integration of rare-earth emitters into scalable, multifunctional photonic circuits for quantum and classical applications, leveraging at best the strong electro-optic, piezoelectric and nonlinear optic control capabilities intrinsically afforded by the x-cut TFLN platform.

## 1 Introduction

Due to its non-centrosymmetric structure, lithium niobate (LN) offers a number of properties, including high second-order nonlinearity and electro-optic effects, which makes it a material of choice for photonic and ntum applications. The demonstration of low-loss $LiNbO_3$ on insulator (LNOI) [1] waveguides has opened up the potential to exploit the properties of this material in small footprint photonic integrated circuits (PICs), mimicking the dense integration of nanophotonic components in silicon on insulator [2, 3]. In recent years notable advances have been reported on LNOI in areas of ultrafast optical modulation [4], nonlinear optics [5], microwave [6] and quantum photonics [7, 8].

With an increasing drive towards realizing 'quantum 2.0' devices utilizing entanglement and superposition, deterministic single-photon sources and memories at telecom wavelengths are the focus of much attention as reliable ways of transmitting, storing and processing quantum information in advanced entanglement-based reconfigurable and large-scale networks ultimately enabled a quantum internet. The PIC platform of LNOI is an attractive proposition for housing such sources, and successful experimental realizations have taken place [9]. Moreover, to further expand the capabilities of this material platform, LNOI PICs have been integrated with other materials and devices such as laser sources [10, 11], superconducting and semiconducting photodetectors [12, 13]. Of particular interest is the inclusion of rare earth ions (REI) within $LiNbO_3$, such as erbium ($Er^{3+}$), due to their useful optical properties and device applications[14, 15, 16]. REI provide transitions in both the optical and microwave domains [17], and their integration into LNOI is attractive for quantum op not to mention tics applications enabling the prospect of a single multi-functional photonic platform. This could provide coherent electro-optic control of optical signals, photon–photon interactions via the large $\chi^{(2)}$ nonlinearity, waveguide-integrated single-photon detection capabilities and, by leveraging the properties of REI, the possibility to implement integrated on-demand single-photon emitters [18] and quantum memories [19], as well as integrated amplifiers as recently demonstrated in silicon nitride [29], on chip,. Experimentally realized examples of REI integration in LNOI can be found for quantum applications [14], light amplification [20, 21, 22] and lasing [15]. The incorporation of $Er^{3+}$ ions is especially useful as it grants these functionalities in the telecom C band, where optical fiber absorption losses are at a minimum.

$Er^{3+}$ ions are typically incorporated into LNOI by applying the smart cut method to traditional Er in-diffused congruent $LiNbO_3$ wafers [23, 24, 25]. This produces a uniformly doped film, and wafers produced by this method are commercially available. However, this approach requires heating the material to 1100 °C [20], approaching the Curie temperature of the crystal (∼ 1200 °C) [26] and has to be implemented in the bulk crystal prior to thin film fabrication and bonding. Furthermore, the high-temperature doping approach does not allow fine spatial control of the $Er^{3+}$ profiles within the LN layer. An alternative to overcome these limitations is ion implantation, which enables deterministic REI incorporation in congruent LNOI at temperatures well below 1000 °C [27] and with extremely high spatial resolution [28], which is ideal for integration in complex PICs. REI can be implanted into predefined regions without affecting the properties of other photonic building blocks integrated on the same chip. Recent developments in focused ion beam (FIB) methods have opened up the possibility of direct writing with spatial resolutions reaching down to few tens of nanometers [28]. This, coupled with deterministic ion implantation and detection, opens up the potential to place single emitters within ready-made waveguide nanostructures suitably engineered for Purcell enhancement and electrical control. Such an approach would avoid the complications of aligning and building the full optical circuitry for resonant enhancement, guidance and control, around randomly distributed pre-existing emitters, as typical for e.g. quantum dots.

In $LiNbO3$, Purcell-enhanced fluorescence has been reported from $Er^{3+}$ (at ∼ 1535 nm [27]) and $Th^{3+}$ (at ∼ 795 nm [30]) ions implanted into nano-waveguides fabricated in z-cut thin films. However, x-cut LNOI is the most useful form for PIC devices, affording in-plane access to the highest electro-optic and nonlinear coefficients [26], as well as enhanced waveguide dispersion engineering capabilities leveraging the material birefringence [31]. The latter feature, together with the strong anisotropy exhibited in the optical and electrical domains by LN waveguides as a result of their crystallographic class and ferroelectric properties (encompassing also pyroelectric, photovoltaic and photorefractive effects), motivate also a thorough investigation of the PL response of Er-implanted x-cut LNOI layers. A deeper understanding of mechanisms that govern emission may provide novel routes for the control of light at the nanoscale. With a view to future quantum applications, the low-temperature behavior of x-cut LNOI is particularly important to understand as many quantum 2.0 devices operate at cryogenic temperatures. In this work we present the integration of $Er^{3+}$ ions into x-cut 500 nm-thick LNOI and study their photoluminescence (PL) intensity and decay properties as a function of temperature and light polarization. The results provide insight on how the internal electric field of LN affects the $Er^{3+}$ emission with pyroelectric behavior observed.

## 2 Results and discussion

### 2.1 Ion-implantation process

Er-doped TFLN samples were prepared for analysis using focused ion beam implantation as described in the experimental section. Although the used implantation system can achieve a minimum feature size as small as 20 nm, in order to ease the optical characterization, we opted for implantation areas of a few hundreds of nm, which can be discerned in the optical microscope images of Figure 1. The writing pattern consisted of larger 100 x 100 µm² squares, accompanied by narrower test stripes of varying widths along the *z* crystallographic axis of the LN film (the smallest being ∽700 nm-wide) written along the *y* axis of the LN film (further details in supplementary information or methods section). The ion implantation was performed using the $^{170}Er$ isotope and implantation doses ranging from $10^{13}$ to $10^{15}$ ions/cm². The implantation energy was set to 75 keV, yielding an Er-ion distribution expected to peak at a depth approximately 25 nm below the exposed surface (Figure 1.a), according to SRIM (Stopping range of Ions in Matter) simulations (Figure 1.e). Figures 1.b-d show optical images obtained by unpolarized brightfield optical microscopy at three different implantation doses. Figure 1.f presents an atomic force microscopy (AFM) image of the interface between an implanted region (surface roughness σ=1.5 nm) and a non-implanted one (σ=0.5) and proves that the process does not significantly affect the surface quality of the film, confirming its potential compatibility with waveguide fabrication techniques on LNOI. After implantation, and prior to photoluminescence (PL) characterization, the samples were annealed in order to let the LN crystal structure recover from any eventual implantation-induced lattice damage and activate the emitters, as described in the experimental methods section. The minimum feature size of the implanted areas did not change after the annealing, thus we concluded that the annealing process, even if relatively slow, did not cause significant solid-state diffusion.

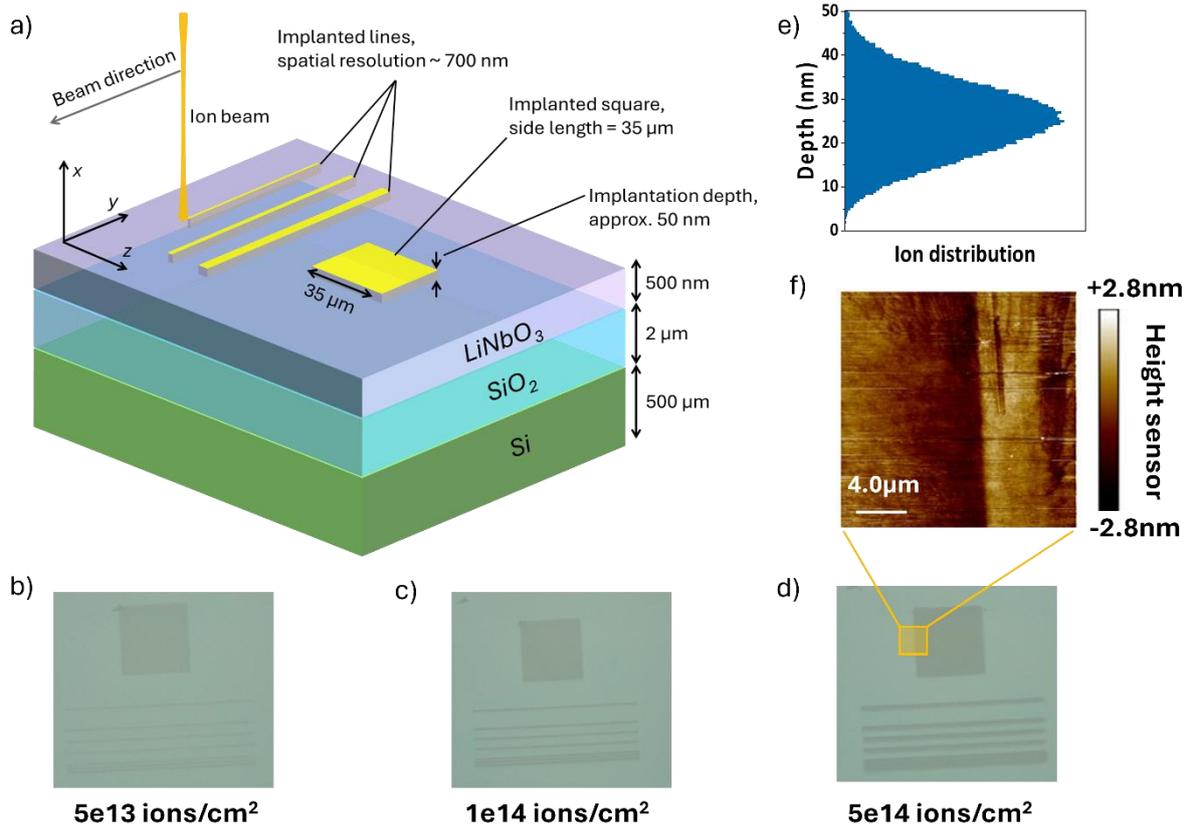

Figure 1. Overview of the experiment. a) Schematic of the implantation process, displaying the Er-implanted features, the structure of the LNOI substrate and the crystallographic axes of the LiNbO$_3$ thin film. b),c) and d) Brightfield optical microscope images of three different implanted regions, characterised by different implantation doses (5 10$^{13}$, 10$^{14}$ and 5 10$^{14}$ ions / cm$^2$, respectively). e) SRIM simulated implanted ion depth distribution, showing that the peak of the implantation depth occurs at approximately 25 nm. f) AFM image collected at the edge of one of the implanted squares (dose 5 10$^{14}$ ions / cm$^2$) confirming that no damage occurred at the level of the LN thin film.

## 2.2 Photoluminescence spectra

PL spectra were obtained from regions implanted with the three different doses, 10$^{13}$, 10$^{14}$, and 10$^{15}$ ions/cm$^2$. The annealed samples were loaded into a cryostat to evaluate the PL at cryogenic temperatures. Figure 2a shows a schematic of the measurement set up (further details in supplementary information). Unless differently specified, the measurements described in what follows were performed with an excitation wavelength of 980 nm and a power of 8 mW from the laser source illuminating a spot size of approximately 1 μm diameter. Microscopic photoluminescence maps across the implanted regions, were recorded by scanning the sample under the excitation beam with piezo-actuators. The results confirmed the deterministic nature of the Er-incorporation process and the sub-micrometric spatial resolutions mentioned above for the smallest implanted features.

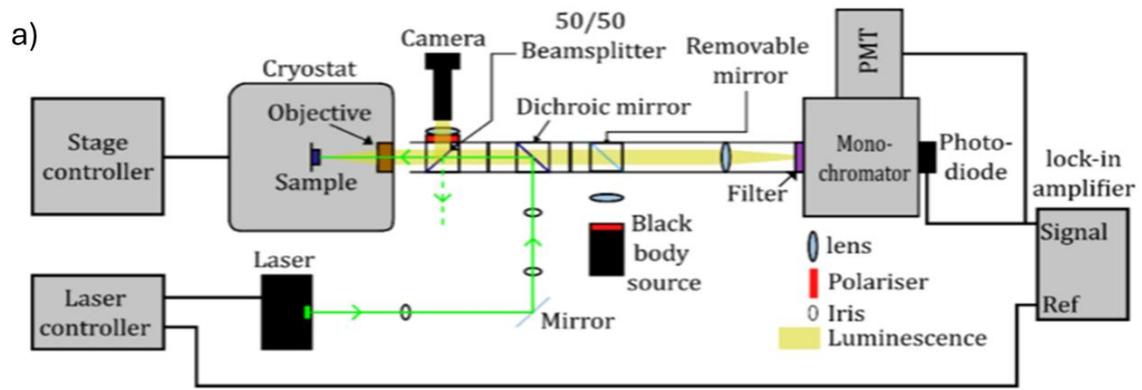

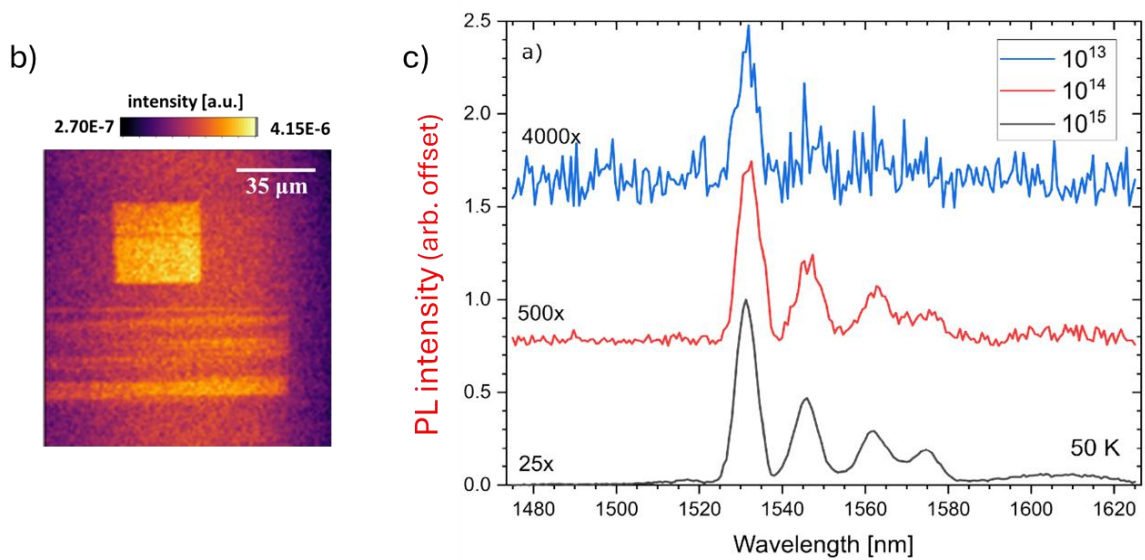

Figure 2: a) Schematic of the experimental setup used to study the photoluminescence of the Er-implanted X-cut TFLN under excitation with 980 nm light . b) Photoluminescence map of one of the large-area Er-implanted regions, taken with a 1200 nm long pass filter in place. c) PL spectra taken at 50 K from sample regions containing $10^{13}$, $10^{14}$, and $10^{15}$ Er ions/cm$^2$.

Further systematic investigations of the spectral and thermal response of the samples were performed focusing on the larger uniformly implanted 100 x 100 µm$^2$ areas, like the one shown in Figure 2b. As can be seen from Figure 2c, the spectra obtained at different irradiation doses show the same features but differ in intensity. The integrated PL intensity was found to be proportional to the ion dose, which is evidence that the dose of $10^{15}$ ions/cm$^2$ had not reached the threshold where the Er$^{3+}$ ion-ion interactions, such as cross-relaxation, have a detrimental effect on the luminescence.

## 2.2 Low-temperature response

Both polarized and unpolarized PL spectra were taken of the implanted regions covering the wavelength region of the $^4I_{13/2} \rightarrow {}^4I_{15/2}$ Er$^{3+}$ transition, at temperatures from room temperature to 5 K. The main results are illustrated by Figure 3.

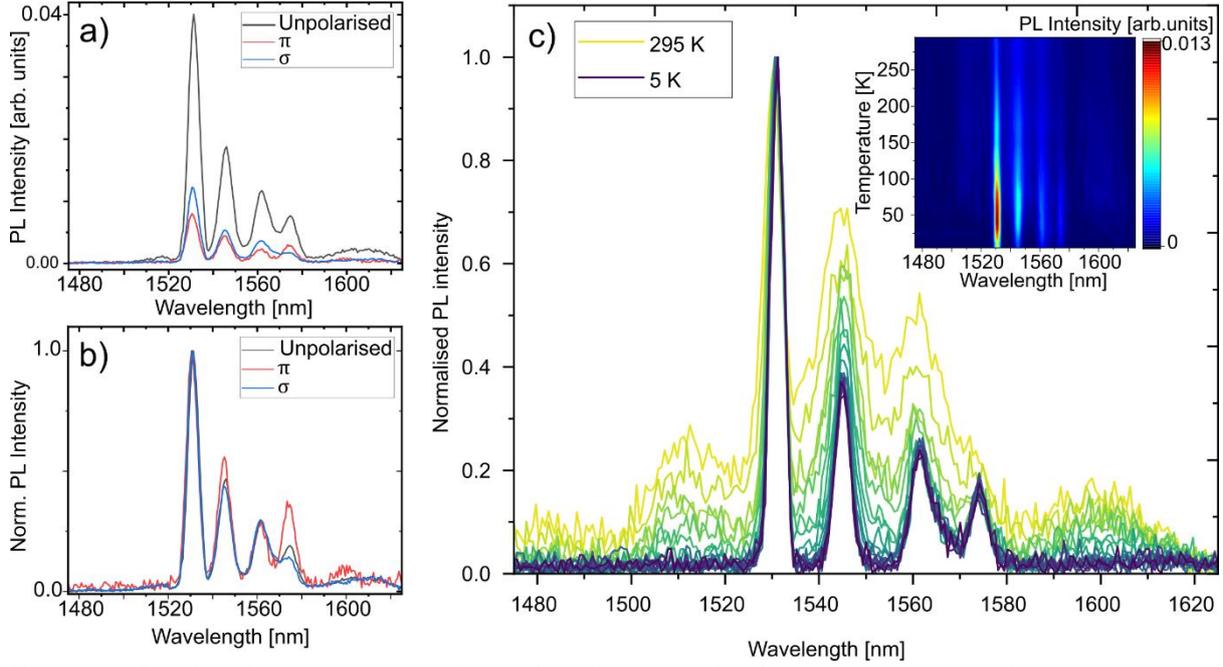

Figure 3: a) Polarized PL spectra taken at 50 K, with a linear polarizer in the detection path either parallel to the crystal z (c)-axis (π), or perpendicular to it (σ). b) A normalized plot of the polarized PL spectra to more easily see the relative peak heights. c) Normalized PL spectra taken at sample temperatures from 295 K (yellow) to 5 K (purple). Inset: heatmap of the absolute spectra over the same temperature range.

The $Er^{3+}$ 4f levels are Stark-split into up to $2J + 1$ non-degenerate Kramer's doublets depending on the local symmetry of the Er ions in the host material, resulting in 8 different energy states in the $^4I_{15/2}$ manifold and 7 in the $^4I_{13/2}$ manifold, labelled $Z_1$ (lowest energy) to $Z_8$ (highest energy) and $Y_1$ to $Y_7$ respectively. Transitions from $Y_1$ to all Z states apart from $Z_8$ were observed in the photoluminescence spectrum at 50 K, as well as the $Y_2 \rightarrow Z_1$ transition, as expected. The energies of these transitions are compared to calculated and experimental literature values for melt-doped $Er:LiNbO_3$ bulk crystals in Table 1. The energies of the transitions in this work are comparable with these literature values indicating the $Er^{3+}$ ions in the ion-implanted LN thin film experience a similar crystal field as those in the melt-doped bulk crystals and are thus occupying similar lattice sites.

| Stark level | $\lambda_{Y_1 \rightarrow Z_n}$ (nm) | Relative E ($cm^{-1}$) | | | | |
| --- | --- | --- | --- | --- | --- | --- |
| | | This work | Exp [34] | Exp [35] | Exp [33] | Exp [36] | Calc [33] |
| $Z_1$ | 1530.9 | 0 | 0 | 0 | 0 | 0 | 0 |
| $Z_2$ | 1545.4 | $60.8 \pm 0.1$ | 63 | 63 | 61 | 67 | 72 |
| $Z_3$ | 1561.7 | $128 \pm 1$ | 129 | 132 | 128 | 127 | 136 |
| $Z_4$ | 1568.5 | $156 \pm 2$ | 152 | 156 | 155 | 167 | 152 |
| $Z_5$ | 1573.9 | $178 \pm 1$ | 185 | 182 | 183 | 191 | 195 |
| $Z_6$ | 1600 | $282 \pm 2$ | 269 | 278 | 270 | * | 301 |
| $Z_7$ | 1613 | $332 \pm 5$ | 353 | 353 | 352 | 367 | 385 |
| $Z_8$ | * | * | 414 | 414 | 415 | 391 | 412 |

Table 1: The wavelengths of the optical transitions to from the $^4I_{13/2}$ $Y_1$ the Stark level to the $^4I_{15/2}$ manifold. The energy of the Stark splitting relative to the Z1 level is shown for this work and for experimental and calculated values reported in the literature for Er-doped bulk LN. The values from this work were obtained by fitting peaks to the observed spectra, at a sample temperature of 50 K. The asterisks for the $Z_8$ Stark level indicate that the emission was too weak to resolve the corresponding peak in the spectra.

The rare earth 4f-4f optical transitions are described by Judd-Ofelt theory [32], which states that the crystal field makes the usually forbidden transitions weakly allowed, and allows to determine the electric and magnetic dipole oscillator strengths which govern the intensity of the transitions. The uniaxial nature of the LN crystal results in different transition strengths for modes polarized parallel and perpendicular to the optical z- (c-) axis of LN, denoted as π-polarized and σ-polarized respectively, in our experimental conditions. The $^4I_{9/2} \leftarrow {}^4I_{15/2}$ excitation at 980 nm was found to be predominantly σ-polarized, which is consistent with literature [33]. The polarized PL spectra for the $^4I_{13/2} \rightarrow {}^4I_{15/2}$ transition can be seen in Figure 2a-b and, which show that all apart from the $Y_1 \rightarrow Z_5$ transition are predominantly σ-polarized, whereas the $Y_1 \rightarrow Z_5$ transition was predominantly π-polarized.

## 2.2 Low-temperature behavior

Figure 3c shows that the FWHM of the $Er^{3+}$ emission peaks narrows monotonically with decreasing temperature, a typical behavior as phononic broadening decreases. More interesting behavior was seen when the spectrally integrated intensity and the PL lifetime were examined as a function of temperature, as shown in Figure 4. The PL decay was fitted using a convolution of the instrument response function and a single exponential decay; the exact functions and procedures used to analyzse the experimental data can be found in the supporting information.

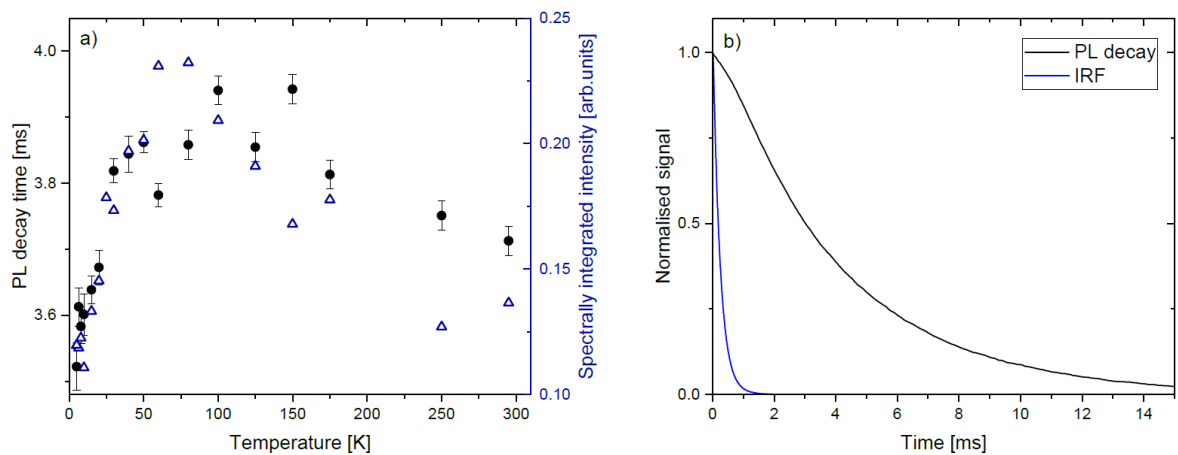

Figure 3: a) The $Er^{3+}$ $^4I_{13/2} \rightarrow {}^4I_{15/2}$ PL decay time (black circles, left axis) and spectrally integrated PL intensity (blue triangles, right axis) as a function of temperature. b) PL decay transient at 80 K showing the detected PL signal and the instrument response function from laser light scatter. Raw data for the other temperatures and fitting details can be found in the supporting information.

Both intensity and decay time increase as the sample is cooled from room temperature to approximately 100 K. This behavior is expected as $k_BT$ decreases below the activation energy of nonradiative recombination processes and for rare-earth ions the intra-atomic 4f-4f multiphonon relaxation rate decreases [37]. However, below 100 K the PL intensity and PL lifetime stabilize on a plateau until approximately 50 K, beyond which they decrease rapidly with further cooling. This decrease amounts to 50% for the PL intensity and 10% for the PL lifetime. This is atypical for the PL of rare-earth ion doped materials and indicative of a host-material specific effect in the LN host matrix.

Although the rich variety of often-linked phononic, electronic and optical effects harbored by LN hinders at this point the unambiguouos attribution of the exact physical mechanism underpinning the observed anomaly in the PL response, the qualitative resemblance of the latter to the reported evolution of the pyroelectric response on LN for decreasing temperatures [38-39] might provide a clue to further insights and connections to the ferroelectric nature of the LN crystal. The ferroelectric nature of LN makes it exhibit a pyroelectric effect, i.e. the generation of a change in its spontaneous polarization ($P_s$= 75 μC/cm$^2$ at room temperature, oriented along the crystal z-axis), which can result in the build-up of a very high electric field in response to a change in temperature. The pyroelectric response, measured as a voltage arising in response to a temperature gradient ∂T, is proportional to the (temperature-dependent) pyroelectric coefficient ($p_3=\partial P_s/\partial T$) divided by the volume heat capacity ($c_p$). In LiNbO$_3$ both quantities decrease as temperature decreases from room temperature, but $c_p$ decreases at a faster rate, resulting in a slight increase of the pyroelectric response. However, below a certain threshold temperature, the lowest-energy optical phonon mode is frozen out. As the latter supports pyroelectricity, $p_3$ drops dramatically, resulting in a maximum in the pyroelectric response occurring at $k_B T \approx 0.2 - 0.3)\ \Omega$, where $\Omega$ is the energy of the lowest-lying optical phonon mode [38]. In LiNbO$_3$ this has been measured as approximately 154 cm$^{-1}$ at low temperatures [39], corresponding to an expected maximum in the pyroelectric response in the 44–66 K range. This is also consistent with the experimental data from Bravina et al. [40], and is also the same temperature range at which we observe the change in luminescence behavior here. In fact, the qualitative evolution of the integrated response of the PL intensity and of the decay time retrieved from our experiments and shown as a function of temperature in Figure 4a are remarkably similar to the pyroelectric response of LiNbO$_3$ as a function of temperature. It is also worth highlighting that the conditions of our experiments are very similar to those typically used to test the pyroelectric response, whereby the sample is exposed to a temporally modulated thermal flux, here heating due to absorption of the 980 nm excitation laser, which can induce a pyroelectric voltage along the optic axis of the LiNbO$_3$. In traditional pyroelectric measurements performed with modulation rates in the 10 - 1000 Hz range, the pyroelectric response is found to be approximately inversely proportional to the modulation frequency of the thermal flux [41]. These conditions compare well with those of our experiments. The intensity of the PL in our sample as a function of temperature was measured at modulation frequencies (with constant 50 % duty cycle) from 2.51 Hz to 204 Hz of the excitation laser, resulting in the data shown in Figure 5. From it, it is apparent that the low-temperature drop-off in intensity is inversely proportional to the modulation frequency, further supporting the hypothesis that it could be due to pyroelectric phenomena.

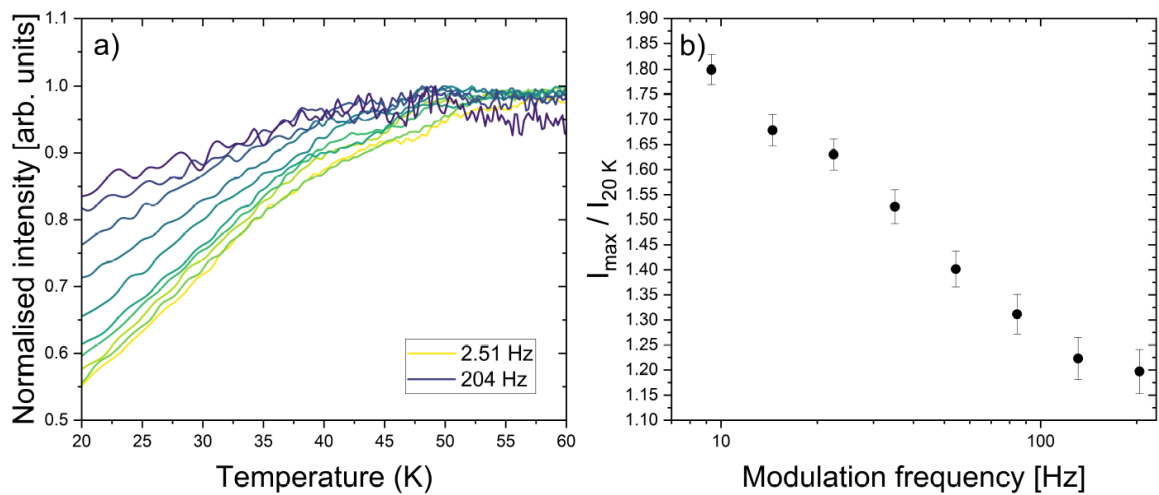

Figure 5: a) Normalized spectrally integrated Er$^{3+}$ 4I$_{13/2}$ → 4I$_{15/2}$ PL intensity as a function of temperature, for excitation laser modulation frequencies logarithmically spaced between 2.51 and 204 Hz. b) Maximum intensity divided by the intensity at 20 K for the same conditions —an indication of the size of the low-temperature drop off in intensity.

For the trends seen in our data to be related to the pyroelectric effect, there must be a physical mechanism for it to affect the luminescence. We speculate this to be the change in ferroelectric features, i.e. the spontaneous polarization, of LN which in turn leads to an electric field, supported by the fundamentally insulating nature of the material. Such an internally arising electric field could either affect the luminescence directly, through the Judd-Ofelt parameters and local field corrections, or could indirectly via the linear electro-optic (Pockels) effect [26]. In any case it should affect the (z-)π-polarized luminescence more than the (x-)σ-polarized one, as its electric field is parallel to that induced by the pyroelectric effect. Polarized spectra taken at different modulation frequencies indicate this is the case for our sample: the π-polarized luminescence was significantly more affected by modulation frequency changes (see supporting information for more details).

One might wonder why this behavior has not been reported in previous studies that have examined low-temperature erbium luminescence in LiNbO$_3$, such as that of Gruber et al. [33]. One difference is that here the Er ions have been implanted instead of incorporated during growth, although the similarity in the Stark splitting of the ground state indicates that the Er ions are occupying the same lattice site so this should not make a difference. Another is that here the LiNbO$_3$ is a 500nm thin film bonded to a 2 µm-thick SiO$_2$ on a Si wafer. Silicon has an absorption coefficient orders of magnitude higher than pure lithium niobate at the excitation wavelength of 980 nm [42, 43]. This leads to much more localized heating from the laser excitation and a larger pyroelectric response than in other studies carried out on large single crystals of LiNbO$_3$.

## 3 Conclusion

In conclusion, we have shown that FIB implantation is an effective way of incorporating Er ions into LN, allowing the utilization of their optical properties within the LN materials platform. The sample was effectively annealed to repair any damage caused by the ion implantation process, analysis of the Stark splitting of the emission suggests the Er$^{3+}$ ions settled to the same lattice sites as in melt-doped bulk LiNbO$_3$.

A systematic temperature-dependent analysis of the luminescence revealed interesting behavior at temperatures below ∼ 50 K, attributed to pyroelectric phenomena in the LN. As Er$^{3+}$ ions in LN are seen as an attractive pathway for quantum memory and information processing devices, performance at these temperatures is critical and therefore this behavior should be taken into account when designing devices — in our case we saw a decrease of 50% in the luminescence intensity cooling from 50 K to 5 K.

To gain a deeper understanding of the physics behind this behavior, further experiments are needed — electrodes can be patterned outside the Er implanted areas on the sample, allowing the application of external electric fields which could elucidate the role of the internal fields generated within the LN. In addition, although they are not as pertinent for on-chip photonic applications, Y- and particularly Z-cut LNOI could be used so that the internal fields in the LN would be along different axes, perpendicular to the electric field of the absorbed and collected light in the case of z-cut.

## 4 Experimental Section

The implantation was performed on commercially available TFLN wafers (NanoLN Inc.) and consisted of a 300 nm-thick X-cut congruent undoped lithium niobate film bonded on 2 µm of thermal oxide grown on 500 µm thick silicon.

*Ion Implantation*: The sample was implanted using the P-NAME facility (Ionoptika, Q-One) [28]
at the University of Manchester, UK, providing isotopic selection and sub-100 nm spatial resolution forEr implantation. The column was configured to implant the 170 Er$^{3+}$ isotope using an 25 kV acceleration voltage (giving in a 75 keV implantation energy) which resulted in a mean implantation depth of 25.4 nm as modelled using the stopping range of ions in matter (SRIM) simulations, see Figure 1.e) for the distribution. SRIM is a Monte-Carlo based package that takes into account the implantation conditions as well as the target material properties to simulate the ion implantation process. The output of interest for this work was the expected depth of the implanted ions. The doses for the series of 100 by 100 µm regions implanted were defined by control of the dwell time of the rastering focused ion beam, with a nominal beam current of 17 pA, to doses of $10^{13}$, $10^{14}$, and $10^{15}$ ions/cm$^2$. All FIB implantation was carried out at room temperature.

*Annealing*: Following implantation the sample was annealed using an Annealsys rapid thermal annealer at 850 ◦C for 60 minutes, with 0.5 ◦C/s heating rate and 0.2 ◦C/s cooling rate, under a constant oxygen flow of 40 sccm.

*Optical and AFM sample inspection*s:

Following ion implantation, images of the samples were acquired using a widefield microscope with a 20× objective lens (Figure 1b–d, corresponding to the three implantation doses). For detailed morphological analysis of the implanted LNOI regions, a Bruker ICON AFM system equipped with an

MESP-RC-V2 tip was employed. To assess the surface roughness between the implanted and non-implanted regions, an area encompassing their interface (indicated by the yellow square in Figure 1d) was scanned in tapping mode. The AFM scan was performed over a 2 μm z-range at a rate of 0.5 Hz, as illustrated in Figure 1f. *Photoluminescence*: Photoluminescence measurements were carried out with the sample mounted in a Montana instruments Cryostation, a closed-cycle cryogen-free cryostat with integrated 100x 0.75 NA microscope objective. The excitation light was provided by a Thorlabs 980 nm laser diode, which was electronically modulated to make use of the lock-in detection method in conjunction with an SRS830 lock-in amplifier. At the laser drive current used for the measurements and with the duty cycle of 50 %, the time averaged power at the sample was approximately 3 mW. The laser spot size on the sample was approximately 1 μm for mapping measurements but for the spectra was defocused to the order of 5 μm to increase the detected signal. A Thorlabs longpass dichroic mirror with a cut-on wavelength of 1180 nm was used to separate the PL and excitation laser beam paths. A FELH1050 1050 nm longpass filter was used before the monochromator to further eliminate any laser light reaching the detector. To collect the PL maps the sample was moved inside the cryostat using an xyz attocube piezo stage. To detect the PL signal for the spectral and time-resolved measurements a NIR-sensitive InP/InGaAs photomultiplier tube was used. This was combined with the lock-in amplifier for the spectra and a Keysight Infiniium S-Series oscilloscope for the time-resolved measurements. The instrument response function (0.23 ms time constant, see Figure 3.b) for a plot and the supporting information for more details) was measured and deconvolved from the time-resolved signal to isolate the contribution of the PL decay to the detected signal.

## Supporting Information

Supporting Information is available from XXXXX from the author.

## Acknowledgments

This work was supported by a joint KTH and University of Manchester international seed project, the EPSRC (grant nos. EP/R025576/1, EP/V001914/1, and EP/R00661X/1 Henry Royce Institute) and by capital investment by the University of Manchester in the UK as well as by the Swedish Research Council (grants nos. VR 2018-04487 and VR 2016-06122 Optical Quantum Sensing Environment) and the Knut and Alice Wallenberg Foundation (WACQT grant no. 2017.099) in Sweden. KTH researchers also gratefully acknowledge access to the MyFab-Albanova Nanofabrication Lab facilities

## References

[1] B. Desiatov, A. Shams-Ansari, M. Zhang, C. Wang, M. Lonˇcar, Optica 2019, 6, 3 380.
[2] D. Zhu, L. Shao, M. Yu, R. Cheng, B. Desiatov, C. Xin, Y. Hu, J. Holzgrafe, S. Ghosh, A. Shams-Ansari, et al., Advances in Optics and Photonics 2021, 13, 2 242.
[3] Z. Li, R. N. Wang, G. Lihachev, J. Zhang, Z. Tan, M. Churaev, N. Kuznetsov, A. Siddharth, M. J.Bereyhi, J. Riemensberger, et al., Nature Communications 2023, 14, 1 4856.
[4] M. Xu, M. He, H. Zhang, J. Jian, Y. Pan, X. Liu, L. Chen, X. Meng, H. Chen, Z. Li, et al., Nature communications 2020, 11, 1 3911.
[5] T. P. McKenna, H. S. Stokowski, V. Ansari, J. Mishra, M. Jankowski, C. J. Sarabalis, J. F. Herrmann,C. Langrock, M. M. Fejer, A. H. Safavi-Naeini, Nature Communications 2022, 13, 1 4532.
[6] L. Shao, M. Yu, S. Maity, N. Sinclair, L. Zheng, C. Chia, A. Shams-Ansari, C. Wang, M. Zhang, K. Lai, et al., Optica 2019, 6, 12 1498.
[7] E. Lomonte, M. A. Wolff, F. Beutel, S. Ferrari, C. Schuck, W. H. Pernice, F. Lenzini, Nature communications 2021, 12, 1 6847.
[8] R. Nehra, R. Sekine, L. Ledezma, Q. Guo, R. M. Gray, A. Roy, A. Marandi, Science 2022, 377, 6612 1333.
[9] S. Aghaeimeibodi, B. Desiatov, J.-H. Kim, C.-M. Lee, M. A. Buyukkaya, A. Karasahin, C. J.

Richardson, R. P. Leavitt, M. Lončar, E. Waks, Applied Physics Letters 2018, 113, 22.
[10] C. Op de Beeck, F. M. Mayor, S. Cuyvers, S. Poelman, J. F. Herrmann, O. Atalar, T. P. McKenna, B. Haq, W. Jiang, J. D. Witmer, et al., Optica 2021, 8, 10 1288.
[11] A. Shams-Ansari, D. Renaud, R. Cheng, L. Shao, L. He, D. Zhu, M. Yu, H. R. Grant, L. Johansson, M. Zhang, et al., Optica 2022, 9, 4 408.
[12] B. Desiatov, M. Lončar, Applied Physics Letters 2019, 115, 12.
[13] X. Sun, Y. Sheng, X. Gao, Y. Liu, F. Ren, Y. Tan, Z. Yang, Y. Jia, F. Chen, Small 2022, 18, 35, 2203532.
[14] S. Dutta, E. A. Goldschmidt, S. Barik, U. Saha, E. Waks, Nano letters 2019, 20, 1 741.
[15] T. Li, K. Wu, M. Cai, Z. Xiao, H. Zhang, C. Li, J. Xiang, Y. Huang, J. Chen, APL Photonics 2021, 6, 10.
[16] R. Bao, Z. Fang, J. Liu, Z. Liu, J. Chen, M. Wang, R. Wu, H. Zhang, Y. Cheng, Laser & Photonics Reviews 2025, 19, 1 2400765.
[17] J. Rochman, T. Xie, J. G. Bartholomew, K. Schwab, A. Faraon, Nature Communications 2023, 14, 1 1153.
[18] L. Yang, S. Wang, M. Shen, J. Xie, H. X. Tang, Nature Communications 2023, 14, 1 1718.
[19] S. Dutta, Y. Zhao, U. Saha, D. Farfurnik, E. A. Goldschmidt, E. Waks, ACS Photonics 2023, 10, 4 1104.
[20] Z. Chen, Q. Xu, K. Zhang, W.-H. Wong, D.-L. Zhang, E. Y.-B. Pun, C. Wang, Optics Letters 2021, 46, 5 1161.
[21] M. Cai, K. Wu, J. Xiang, Z. Xiao, T. Li, C. Li, J. Chen, IEEE Journal of Selected Topics in Quantum Electronics 2021, 28, 3: Hybrid Integration for Silicon Photonics 1.
[22] Y. Zhang, Q. Luo, S. Wang, D. Zheng, S. Liu, H. Liu, F. Bo, Y. Kong, J. Xu, Optics Letters 2023, 48, 7 1810.
[23] R. Brinkmann, W. Sohler, H. Suche, Electronics Letters 1991, 27, 5 415.
[24] I. Baumann, R. Brinkmann, M. Dinand, W. Sohler, S. Westenhofer, IEEE journal of quantum electronics 1996, 32, 9 1695.
[25] D. Brüske, S. Suntsov, C. E. Rüter, D. Kip, Optics Express 2017, 25, 23 29374.
[26] R. S. Weis, T. K. Gaylord, Applied Physics A Solids and Surfaces 1985, 37, 4 191.
[27] S. Wang, L. Yang, R. Cheng, Y. Xu, M. Shen, R. L. Cone, C. W. Thiel, H. X. Tang, Applied Physics Letters 2020, 116, 15.
[28] M. Adshead, M. Coke, G. Aresta, A. Bellew, M. Lagator, K. Li, Y. Cui, R. Cai, A. Almutawa, S. Haigh, K. Moore, N. Lockyer, C. Gourlay, R. Curry, Advanced Engineering Materials 2023.
[29] Y. Liu, Z. Qiu, X. Ji, A. Lukashchuk, J. He, J. Riemensberger, M. Hafermann, R. N. Wang, J. Liu, C. Ronning, et al., Science 2022, 376, 6599 1309.
[30] D. Pak, A. Nandi, M. Titze, E. S. Bielejec, H. Alaeian, M. Hosseini, Communications Physics 2022, 5, 1 89.
[31] J. Schollhammer, M. A. Baghban, K. Gallo, Optics letters 2017, 42, 18 3578.
[32] B. R. Judd, Physical Review 1962, 127, 3 750.
[33] J. B. Gruber, D. K. Sardar, R. M. Yow, B. Zandi, E. P. Kokanyan, Physical Review B - Condensed Matter and Materials Physics 2004, 69, 19 1.
[34] D. M. Milori, I. J. Moraes, A. C. Hernandes, R. R. De Souza, M. S. Li, M. C. Terrile, G. E. Barberis, Physical Review B 1995, 51, 5 3206.
[35] V. T. Gabrielyan, A. A. Kaminskii, L. Li, Physica Status Solidi (a) 1970, 3, 1 K37.
[36] G. Dominiak-Dzik, S. Go lab, I. Pracka, W. Ryba-Romanowski, Applied Physics A 1994, 58 551.
[37] D. Yu, J. Ballato, R. E. Riman, Journal of Physical Chemistry C 2016, 120, 18 9958.
[38] A. M. Glass, M. E. Lines, Physical Review B 1976, 13, 1 180.
[39] A. R. Zanatta, Results in Physics 2023, 47, February 106380.
[40] S. L. Bravina, A. N. Morozovska, N. V. Morozovsky, Y. A. Skryshevsky, Ferroelectrics 2004, 298 31.
[41] S. L. Bravina, N. V. Morozovsky, A. N. Morozovska, S. Gille, J. P. Salvestrini, M. D. Fontana, Ferroelectrics 2007, 353, 1 PART 3 202.
[42] K. L. Sweeney, L. E. Halliburton, Applied Physics Letters 1983, 43, 4 336.
[43] M. A. Green, M. J. Keevers, Progress in Photovoltaics: Research and Applications 1995, 3, 3 189.